\documentclass[conference,review]{IEEEtran}

\IEEEoverridecommandlockouts
\usepackage{tikz}

\newcommand{\fullbullet}{%
  \tikz[baseline=-0.6ex]\fill (0,0) circle (0.55ex);%
}

\newcommand{\emptybullet}{%
  \tikz[baseline=-0.6ex]\draw (0,0) circle (0.55ex);%
}

\newcommand{\halfbullet}{%
  \tikz[baseline=-0.6ex]{
    \draw (0,0) circle (0.55ex);
    \fill (0,-0.55ex) arc[start angle=-90,end angle=90,radius=0.55ex] -- cycle;
  }%
}

\usepackage{amsmath,amssymb,amsfonts}
\usepackage{algorithmic}
\usepackage{graphicx}
\usepackage{textcomp}
\usepackage{xcolor}
\def\BibTeX{{\rm B\kern-.05em{\sc i\kern-.025em b}\kern-.08em
    T\kern-.1667em\lower.7ex\hbox{E}\kern-.125emX}}

\usepackage{booktabs}
\usepackage{array}
\usepackage{makecell}
\usepackage{tabularx}
\usepackage{xcolor}
\usepackage{makecell}
\usepackage{placeins}
\usepackage{colortbl}
\usepackage[T1]{fontenc}
\usepackage{textcomp}
\usepackage{amssymb}
\usepackage{caption}
\usepackage{amsmath,amsfonts}
\usepackage{algorithm}
\usepackage{soul}
\usepackage{adjustbox}
\usepackage{svg}
\usepackage{array}
\usepackage{multirow}
\usepackage[acronym,nopostdot,nogroupskip,toc,nonumberlist]{glossaries}
\usepackage{glossary-longragged} 
\setglossarystyle{tree}
\usepackage{longtable}
\usepackage{rotating}
\usepackage[caption=false,font=normalsize,labelfont=sf,textfont=sf]{subfig}
\usepackage{textcomp}
\usepackage{pifont} 
\usepackage{stfloats}
\usepackage{url}
\usepackage{verbatim}
\usepackage{graphicx}
\usepackage{multirow}
\ifCLASSOPTIONcompsoc
  \usepackage[nocompress]{cite}
\else
  \usepackage{cite}
\fi
\usepackage{xurl}

\newcolumntype{L}[1]{>{\raggedright\arraybackslash}p{#1}}

\begin{document}

\title{\Large \bf FuseChain: Runtime Evidence Reconstruction for Software Supply-Chain Attacks}

\author{
\IEEEauthorblockN{
Zhuoran Tan\IEEEauthorrefmark{1},
Yutian Tang,
Jeremy Singer,
Christos Anagnostopoulos,
Ke Xiao
}
\IEEEauthorblockA{
School of Computing Science, University of Glasgow, Glasgow, United Kingdom\\
\{z.tan.1, Yutian.Tang, jeremy.singer, Christos.Anagnostopoulos, k.xiao.1\}@glasgow.ac.uk
}
\thanks{\IEEEauthorrefmark{1}Corresponding author.}
}

\maketitle

\begin{abstract}
Software supply-chain (SSC) attacks are increasingly multi-stage, cross-source, and temporally distributed. A single attack campaign may leave weak and fragmented traces across multi-source telemetry that captures different granularities and perspectives of runtime behavior. Existing runtime detection systems often analyze these sources independently, making it difficult to identify low-frequency attack evidence or reconstruct the temporal context in which it appears.
We present \textsc{FuseChain}, a runtime detection framework that represents multi-source software supply-chain telemetry as a temporal heterogeneous provenance graph over a unified event-time axis. By aligning package/runtime traces, process events, network telemetry, DNS/HTTP metadata, and security alerts on a unified temporal graph, FuseChain captures cross-source dependencies and sparse attack evidence that may be ambiguous within any individual source. It learns anomaly-centric temporal representations from benign-prefix telemetry and performs deployable attack-stage reconstruction through a lightweight decoder on top of a frozen anomaly backbone.
Our experiments show that jointly optimizing anomaly detection and stage prediction is ineffective under sparse and imbalanced runtime supply-chain telemetry. Across seven SSC attack scenarios, FuseChain improves deployable stage reconstruction from 0.369 to 0.881 Stage Recall@500 with a frozen-backbone decoder, while adaptive retrieval further increases observable-stage recall from 0.524 to 0.655 without modifying the detector. These results highlight the deployable value of decoupling runtime SSC anomaly detection from downstream attack-stage interpretation.

\end{abstract}


\section{Introduction}
SSC attacks (Solarwind \cite{alkhadra_solar_2021}, XZ utils \cite{11025592}, 3CX \cite{wardle2023macingsense}) compromise software artifacts, dependencies, build pipelines, or runtime components to reach downstream users. Unlike traditional malware execution, modern supply-chain attacks are often multi-stage and cross-source. Initial delivery may appear in package or HTTP logs, execution in process telemetry, persistence in system events, command-and-control in DNS or TLS flows, and exfiltration in network traffic~\cite{Duan2020TowardsMS}.

This creates a runtime detection challenge. Individual telemetry sources often contain only weak or ambiguous evidence. A network connection may appear benign without process context; a process execution may appear normal without package provenance; and an Intrusion Detection System (IDS) alert may not reveal the surrounding dependency, execution, and file-access context~\cite{Cheng2023KairosPI, 9789878}. Effective runtime SSC defense therefore requires temporal reasoning over heterogeneous evidence distributed across multiple sources.

Although provenance-based intrusion detection has shown the value of graph modeling, most existing systems target enterprise APT settings or host-centric audit streams~\cite{Cheng2023KairosPI, jian2025, li2024nodlink, 294545}. Runtime SSC telemetry is broader and sparser: evidence may span multi-source runtime telemetries, while different attack stages may be visible only through low-volume signals from different sources~\cite{tan_2026_synthchain}.

To address these challenges, we propose \textbf{FuseChain}, a runtime SSC detection framework that aligns heterogeneous telemetry into typed temporal events over a unified event-time axis. FuseChain learns anomaly-centric graph representations from benign-prefix telemetry and reconstructs observable attack stages through a lightweight downstream decoder on top of a frozen anomaly backbone. This design separates behavioral anomaly learning from deployable attack-stage interpretation, avoiding the interference caused by sparse and imbalanced stage supervision.

A key empirical finding is that attack-stage interpretation is more effective when decoupled from anomaly detection. Although stage supervision is appealing, sparse and imbalanced stage labels can interfere with anomaly-centric representation learning. FuseChain therefore freezes the anomaly backbone and trains a lightweight downstream decoder, improving deployable stage reconstruction while preserving detection quality.
In summary, this work makes the following contributions:
\begin{itemize}
    \item We propose FuseChain, a runtime SSC detection framework that aligns heterogeneous telemetry sources into a unified temporal provenance graph for cross-source evidence modeling.
    \item We design a self-supervised temporal graph anomaly detector that learns anomaly-centric representations from benign-prefix telemetry without relying on attack-stage labels during detector training.
    \item We demonstrate that deployable attack-stage reconstruction is more effective when treated as a downstream interpretation task over a frozen anomaly backbone, rather than as a jointly optimized objective.
    \item We introduce IOC traceability and observable-stage reconstruction for analyst-facing attack-chain recovery.
    \item We evaluate FuseChain on seven SSC attack scenarios in SynthChain~\cite{tan_2026_synthchain}, showing improved full-scenario detection and reconstruction, a deployable Stage Recall@500 increase from 0.369 to 0.881, and an adaptive retrieval gain from 0.524 to 0.655 observable-stage recall.
\end{itemize}

Our implementation and experiment scripts are available in an anonymous artifact repository.\footnote{\url{https://anonymous.4open.science/r/graphchain-detection-DF9C}}

\section{Background and Threat Model}

\subsection{Software Supply Chain Attack Stages}
SSC attacks are rarely single-step compromises. They typically unfold through operational stages that may span multiple systems and telemetry sources. Following common attack-lifecycle models~\cite{strom2020attack}, FuseChain considers seven observable stage categories: \emph{Resource Development}, \emph{Initial Access/Delivery}, \emph{Execution}, \emph{Persistence/Privilege Escalation/Defense Evasion}, \emph{Discovery/Collection}, \emph{Command and Control (C2)}, and \emph{Exfiltration/Impact}.

Unlike traditional endpoint attacks, evidence for these stages is often fragmented across package-management systems, runtime environments, operating-system telemetry, and network infrastructure. For example, payload delivery may appear in HTTP logs, execution may be visible only through process telemetry, while C2 activity may be observable exclusively through DNS or TLS metadata.

Consequently, reconstructing attack progression requires temporal correlation across heterogeneous telemetry sources rather than isolated analysis of individual logs.

\subsection{Runtime Multi-Source Telemetry}

FuseChain assumes access to multi-layer runtime telemetry sources described above, covering package, host, network, cloud, and alert-level observations. These sources provide complementary visibility into software execution, dependency behavior, network communication, and security alerts, but differ in structure, semantics, and granularity.

FuseChain does not assume IOC annotations or attack-stage labels for anomaly-backbone training; the detailed label-use protocol is described in Section~\ref{sec:methodology}.

\subsection{Threat model}

\textbf{Attacker}: We consider an attacker who compromises SSC components, including software packages, dependencies, build artifacts, update mechanisms, or credentials. The attacker seeks to execute malicious actions while remaining inconspicuous within any individual telemetry source.

\textbf{Defender}: The defender observes heterogeneous runtime telemetry but does not rely on IOC annotations or attack-stage labels when training the anomaly detector. FuseChain surfaces high-scoring anomalous events, localizes them into suspicious evidence clusters, and reconstructs observable attack stages to support analyst investigation. Final attack confirmation relies on telemetry context, alerts, threat intelligence, IOC matches, or analyst validation.

\textbf{Scope}: FuseChain focuses on runtime detection after deployment. It complements rather than replaces static package analysis, dependency scanning, or software composition analysis tools.

\section{Problem Formulation}
\label{sec:problem formula}

We model runtime software supply-chain telemetry as a time-ordered sequence of typed interaction events:
$\mathcal{S}=\{e_i\}_{i=1}^{N}$,
where each event is represented as
\[
e_i = (\hat{t}_i, u_i, r_i, v_i, a_i, \sigma_i),
\]
where $\hat{t}_i$ denotes the unified event time, $u_i$ and $v_i$ are typed source and destination entities, $r_i$ is the relation type, and $a_i$ contains multi-attribute evidence extracted from the raw log record, and $\sigma_i$ identifies the telemetry source. If a real timestamp is available, $\hat{t}_i$ is the observed timestamp; otherwise, event order is used as pseudo-time. This definition supports both timestamped logs and ordered traces without assuming that all sources share the same native clock or granularity.

The observed temporal provenance graph at event time $\hat{t}$ is
\[
G_o(\hat{t}) = (V, E_o(\hat{t})),
\]
where each observed event induces a typed edge in \(E_o\). Optionally, we construct a derived dependency edge set \(E_c(\hat{t})\) using deterministic temporal rules, yielding
\[
G(\hat{t}) = (V, E_o(\hat{t}) \cup E_c(\hat{t})).
\]
The derived edges are not additional observations or attack labels; they are rule-based temporal dependency augmentations used only to improve graph context, and their impact is evaluated through ablation.

FuseChain addresses three linked tasks:
\begin{enumerate}
    \item \textbf{Event-level anomaly ranking.} Given a temporal graph stream, the detector assigns each observed event $e_i$ an anomaly score $s_i$, where higher scores indicate lower temporal predictability under the learned normal graph dynamics.
    \item \textbf{Suspicious subgraph localization.} High-scoring events are grouped into analyst-facing evidence clusters using temporal proximity, shared entities, and provenance relationships. This step provides contextual explanations rather than isolated alerts.
    \item \textbf{Observable attack-stage reconstruction.} Let $S_{\mathrm{obs}}$ denote the set of attack stages that have IOC-backed evidence in the processed telemetry stream. Given a retrieved evidence set $R_K$, FuseChain reconstructs the subset of observable stages supported by the retrieved events and their temporal ordering. The objective is not to recover a complete semantic kill chain, but to reconstruct the attack stages that are observable under the available runtime telemetry.
\end{enumerate}

In black-box reconstruction, stage assignments are derived from ground-truth (GT) IOC-stage mappings to evaluate the quality of anomaly ranking. In deployable reconstruction, stage assignments are predicted by a downstream stage decoder, reflecting the setting in which GT stage labels are unavailable during operation.

\section{Methodology}
\label{sec:methodology}

\begin{figure*}[!htbp]
    \centering
    \footnotesize
    \captionsetup{justification=centering}
    \includegraphics[scale=0.315]{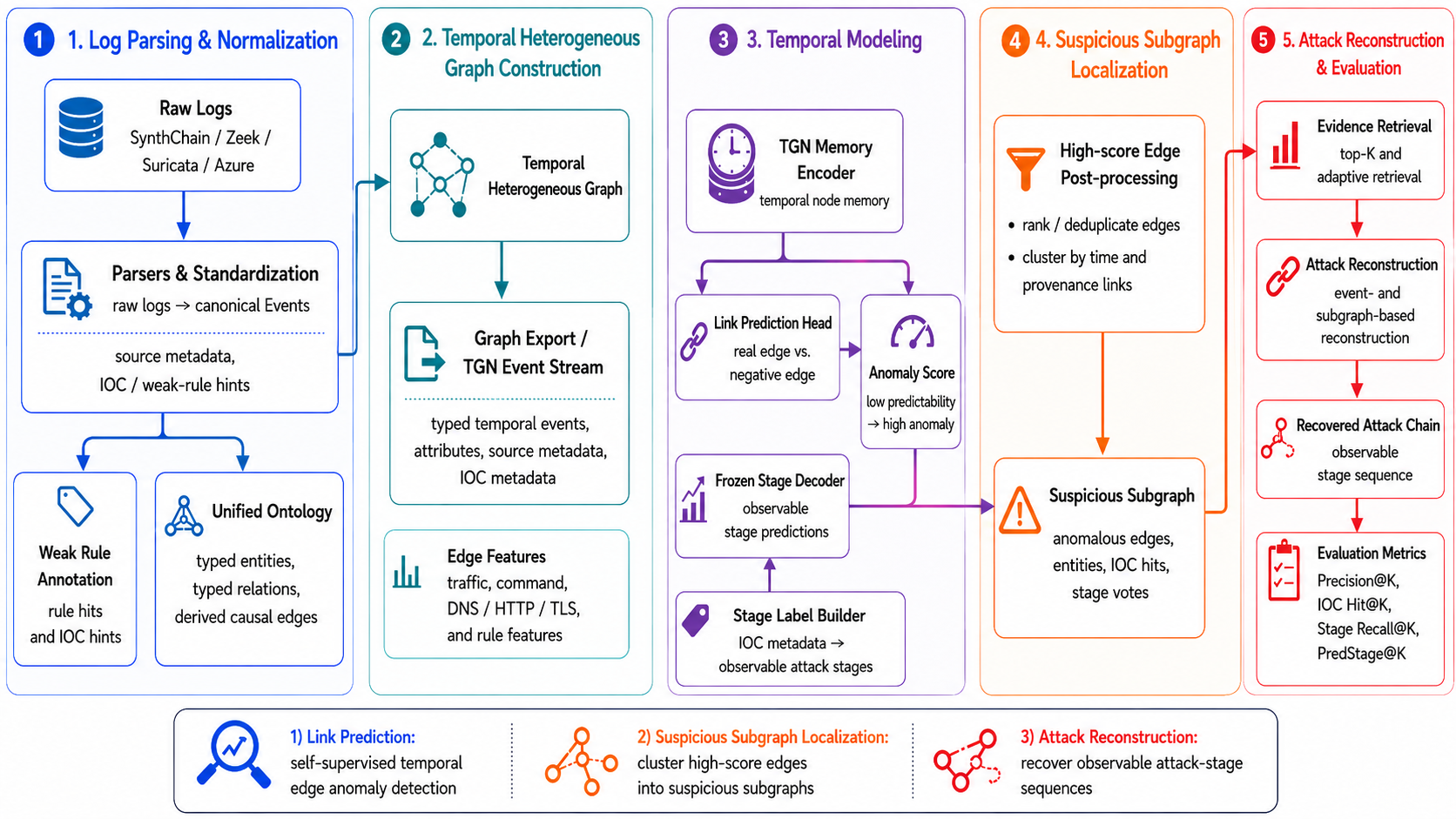}
    \caption{FuseChain Framework}
    \caption*{\small Core pipeline from raw logs to anomalous edges, anomalous subgraphs, and attack-chain reconstruction.}
    \label{fig:fusechain-framework}
\end{figure*}

Figure~\ref{fig:fusechain-framework} gives an overview of \textsc{FuseChain}. The framework takes heterogeneous runtime telemetry as input. It first normalizes raw records into typed temporal events with source-level metadata and weak-rule annotations, then constructs a temporal heterogeneous provenance graph over a unified event-time axis. On top of this graph, FuseChain learns anomaly-centric representations through self-supervised temporal link prediction and assigns anomaly scores to observed events. High-scoring events are then post-processed into suspicious evidence clusters, which are used for attack-stage reconstruction and evaluation.

\textbf{Training and label-use protocol}. FuseChain separates anomaly-backbone training from downstream interpretation and evaluation. The anomaly backbone is trained self-supervised on a chronological historical prefix using temporal link prediction, without using IOC annotations or attack-stage labels as input features or supervision. The prefix is not assumed to be perfectly clean and may contain unlabeled or IOC-bearing events, reflecting realistic historical telemetry. IOC annotations are retained only as external metadata for traceability and evaluation. When deployable stage reconstruction is evaluated, attack-stage labels are used only to train a lightweight decoder on top of the frozen anomaly backbone; they do not update the anomaly detector.

This design separates three concerns: multi-source temporal evidence modeling, self-supervised anomaly detection, and deployable attack-stage interpretation. The following subsections describe each component in order.

\subsection{Multi-Source Temporal Provenance Graph}

FuseChain first normalizes heterogeneous security telemetry into the canonical temporal event schema in Section~\ref{sec:problem formula}, mapping each raw record to typed source and destination entities, relation, attributes, timestamp or pseudo-time, and source identifier.

FuseChain maps heterogeneous telemetry into a compact ontology with typed entities, observed relations, and optional derived causal edges; Table~\ref{tab:ontology-summary} summarizes the schema. 

\begin{table}[t]
\centering
\caption{Summary of the FuseChain ontology. Full node/edge schemas and parser mappings are included in the supplementary artifact.}
\label{tab:ontology-summary}
\small
\begin{tabularx}{\columnwidth}{@{}l>{\raggedright\arraybackslash}X@{}}
\toprule
Category & Types \\
\midrule
Node types &
PKG, PROC, CMD, FILE, NET, HOST, USER, CRED, ARTIFACT, ALERT, SYSCALL \\
Observed edges &
DEPEND, LOAD, EXEC, READ, WRITE, DELETE, CONNECT, DNS\_QUERY, RESOLVE,
REDIRECT, EXFILTRATE, USES\_CRED, ALERT\_ON, INJECT \\
Derived edges &
CAUSE \\
\bottomrule
\end{tabularx}
\end{table}

The full ontology, source-specific parser mappings, and schema-level exploitation coverage are provided in the supplementary artifact.

The graph is heterogeneous in both node types and relation types. However, the TGN model consumes an event stream, so each heterogeneous interaction is flattened into a temporal event while preserving relation type and attribute information. For each event, FuseChain constructs a message vector:
\[
m\_i = [\phi(a_i) , || , \mathrm{Emb}(r_i)],
\]
where $\phi(a_i)$ projects event attributes, $\mathrm{Emb}(r_i)$ embeds the relation type, and $||$ denotes concatenation.

By projecting heterogeneous telemetry into a common graph abstraction, FuseChain enables temporal reasoning across sources that would otherwise remain isolated.

\subsection{Self-Supervised Temporal Link Prediction}

To model normal runtime behavior, FuseChain employs a relation-aware TGN that learns the evolution of the provenance graph in a self-supervised manner.

Each node maintains a temporal memory state that summarizes its historical interactions. Given an observed interaction:
$
(u_i, r_i, \tilde{v}_i, \hat{t}_i).
$
the model updates the corresponding node memories and computes a representation that captures both structural and temporal context.

Training is formulated as a temporal link-prediction task. Observed interactions are treated as positive samples, while unobserved interactions generated through negative sampling serve as negative examples. The model parameters are optimized using binary cross-entropy loss.

After training, the anomaly score of an observed event is defined as:
$
s_i = -\log \sigma(f_\theta(u_i,r_i,v_i,\hat{t}_i,m_i)),
$
where $f_\theta$ is the TGN link predictor and $\sigma$ is the sigmoid function. A higher score indicates that the event is less predictable under the learned normal dynamics. Events that are difficult to explain under the learned temporal dynamics receive higher anomaly scores and are therefore ranked as more suspicious.

Unlike prior approaches that incorporate attack semantics directly into representation learning~\cite{10.1145/3705304, 10646725}, FuseChain intentionally learns anomaly-centric representations without stage supervision. This design isolates behavioral modeling from downstream interpretation tasks.

\subsection{Chronological Historical-Prefix Training}

Historical runtime telemetry cannot be assumed to be perfectly clean: if attack-related interactions appear in the training stream, a self-supervised model may partially absorb them into the normal behavioral distribution. FuseChain therefore uses a chronological historical-prefix protocol. The temporal stream is split into an earlier training prefix and a later evaluation tail; the anomaly backbone is trained only on the prefix, and anomaly detection, evidence localization, and reconstruction are performed on the tail.

As described in the label-use protocol, IOC annotations and attack-stage labels are not used as input features or supervision for anomaly-backbone training. This setting reflects realistic deployment, where the historical prefix is expected to be predominantly benign but may be weakly contaminated.

\subsection{Negative Sampling for Self-Supervised Temporal Learning}

FuseChain trains the anomaly backbone with self-supervised temporal link (SSTL) prediction. For each observed interaction $e_i=(\hat{t}_i,u_i,r_i,v_i,a_i,\sigma_i)$, we construct a corrupted interaction by keeping $u_i$, $r_i$, $\hat{t}_i$, and the event message fixed, while replacing the destination entity $v_i$ with a sampled destination $\tilde{v}_i$. The reported primary configuration uses graph-wide random negative sampling, where $\tilde{v}_i$ is sampled uniformly from the global entity set excluding $v_i$. We use graph-wide random negatives as the primary setting because they avoid relation- or window-specific heuristics and remain stable across scenarios.

We also evaluated time-local hard negatives, where $\tilde{v}_i$ is sampled from destinations appearing with the same relation type within a local training-prefix time window. This strategy can improve anomaly ranking in some scenarios but did not yield consistent gains across all scenarios. We therefore report graph-wide random negative sampling as the primary setting and treat time-local hard negatives as a sensitivity ablation.

\subsection{Decoupled Stage Interpretation via a Frozen Decoder}

Attack-stage prediction is formulated as a downstream interpretation task rather than a joint optimization objective.

Following self-supervised anomaly training, the temporal graph backbone is frozen:
$ \theta \leftarrow \text{fixed}$. For each event, the learned representation $h_i=f_\theta(e_i)$ is extracted and used as input to a lightweight stage decoder: $\hat y_i^{stage}=g_\phi(h_i)$. The decoder is trained using standard cross-entropy loss: $\mathcal{L}_{stage}=CE(\hat y_i^{stage},y_i^{stage})$, where $y_i^{stage}$ denotes the observable attack-stage label when available.

This design is motivated by the observation that stage labels are sparse, highly imbalanced, and often incomplete. Joint optimization causes stage gradients to interfere with anomaly-centric representation learning, leading to degraded detection and reconstruction performance. By freezing the anomaly backbone, FuseChain preserves behavioral representations while enabling stage-level interpretation through a lightweight downstream classifier.

\subsection{IOC Traceability}

To support forensic analysis, FuseChain preserves explicit links between graph interactions and their originating telemetry records. Each event stores its \textit{source file identifier, original row index, and IOC annotations}. We maintain two complementary labels: $y_i^{ioc}$ and $y_i^{line}$. The former captures value-level IOC matches, while the latter identifies exact source records associated with known attack evidence. This mechanism enables analysts to trace suspicious graph interactions back to the underlying telemetry that generated them.

\subsection{Attack-Chain Reconstruction}

FuseChain reconstructs attack evidence from anomalous events through a two-stage retrieval and interpretation process.

\paragraph{\textbf{Black-box Reconstruction}}
This stage evaluates whether anomaly ranking successfully surfaces attack-relevant evidence. Events are first ranked according to anomaly score, and the top-(K) interactions are selected: $\mathcal{R}_K=\text{TopK}(s_i)$.

GT IOC-stage mappings are then used to assign observable attack stages to selected events. This setting isolates the quality of anomaly ranking from the quality of stage prediction and provides an upper bound on reconstruction performance.

\paragraph{\textbf{Deployable Reconstruction}}
In practical deployments, GT stage labels are unavailable. FuseChain therefore replaces black-box stage assignments with predictions generated by the frozen stage decoder.

The resulting reconstruction pipeline evaluates whether suspicious events identified by anomaly detection can be translated into meaningful attack-stage interpretations under realistic operational conditions.

\subsection{Adaptive Retrieval for Reconstruction}

From experiments, we find high anomaly scores are often concentrated within repeated benign structures, causing analyst-visible rankings to become saturated by redundant evidence. Consequently, reconstruction quality may be limited by retrieval diversity rather than anomaly-detection accuracy.

To address this issue, FuseChain evaluates several retrieval strategies on top of the same anomaly ranking. They are:
\begin{enumerate}
    \item The default strategy, \textsc{by\_k}, simply selects the global top-$K$ events by anomaly score. 
    \item \textsc{PairDedupe} collapses repeated endpoint patterns by keeping the highest-scoring event for each $(r,u,v)$ tuple, reducing cases where the same process--object interaction repeatedly occupies the ranking. 
    \item \textsc{SourceQuota} enforces minimum exposure for selected telemetry sources, preventing low-volume but security-relevant logs from being completely suppressed by high-volume sources. 
    \item \textsc{GroupCap} prevents one repeated event pattern from dominating the top-$K$ list. For example, it limits how many events with the same source edge type key can be selected, while still allowing several high-scoring events from that pattern to remain.
    \item Finally, \textsc{Adaptive} first expands the candidate pool to $K'>K$ and then applies hot-group suppression only to groups whose frequency exceeds a threshold. This preserves the original score ordering as much as possible while improving structural diversity in the final top-$K$ evidence set. 
\end{enumerate}

All retrieval strategies operate exclusively on ranked outputs and do not modify the underlying anomaly detector. By increasing structural diversity among retrieved events, adaptive retrieval improves attack-stage coverage while preserving the anomaly-learning process.

\section{Evaluation}

We organize the evaluation around four research questions:
\begin{itemize}
    \item \textbf{RQ1: Detection and localization.} Can FuseChain rank IOC-containing runtime events above benign activity under sparse and imbalanced SSC telemetry?
    \item \textbf{RQ2: Multi-source and temporal modeling.} How does FuseChain compare with single-source baseline and static multi-source graph baselines under same temporal split?
    \item \textbf{RQ3: Deployable attack-stage reconstruction.} Can suspicious events be translated into observable attack-stage sequences without relying on GT stage annotations during deployment?
    \item \textbf{RQ4: Robustness and retrieval effects.} How do retrieval diversity, stage-decoder design, cross-scenario generalization, and weak-rule updates affect reconstruction quality?
\end{itemize}

The primary experiments use graph-wide random negative sampling, where the corrupted destination is sampled uniformly from the global entity set excluding the true destination. 
All experiments were run on a Linux server with Ubuntu 22.04.5 LTS, 16 Intel Xeon Gold 5222 CPU threads, 251~GiB RAM, and one NVIDIA GeForce RTX 3090 GPU with 24~GiB memory. The software environment used Python 3.10.13 and PyTorch 2.8.0 with CUDA support.

\subsection{Dataset and Setup}
\label{sec:dataset}

\paragraph{Dataset}

We evaluate FuseChain on seven SSC attack scenarios from SynthChain. This dataset contains benign background activity, sparse attack-related evidence, and multi-stage SSC behaviors, making it suitable for evaluating deployment-like runtime detection and reconstruction.

The telemetry coverage is scenario-dependent. Across the dataset, SynthChain includes heterogeneous runtime event records derived from process and system activity, network flows, DNS and HTTP/TLS metadata, and security alerts, but not every scenario contains all source types. FuseChain therefore operates on the collected event records and normalizes them into a unified temporal interaction stream before constructing the temporal provenance graph. We use a chronological split (70/30) in which the prefix is used for self-supervised anomaly-backbone training and the tail is used for anomaly detection, evidence localization, and reconstruction.

\begin{table}[!htbp]
\centering
\caption{Semantic and observable stage coverage after IOC refinement. Observable stages are semantic stages with IOC-backed evidence in the processed telemetry stream.}
\label{tab:stage-coverage}
\begin{tabular}{lccc}
\toprule
Scenario & Semantic & Observable & Unobserved \\
\midrule
sc1 & 7 & 4 & 3 \\
sc2 & 7 & 6 & 1 \\
sc3 & 7 & 4 & 3 \\
sc4 & 7 & 6 & 1 \\
sc5 & 6 & 4 & 2 \\
sc6 & 7 & 6 & 1 \\
sc7 & 7 & 6 & 1 \\
\midrule
Mean & 6.9 & 5.1 & 1.9 \\
\bottomrule
\end{tabular}
\end{table}

SynthChain provides IOC metadata with indicator values and source-line evidence. We refine the IOC annotations by adding missing runtime evidence types only when directly supported by existing telemetry records. This refinement does not modify graph construction, model training, or anomaly scores, but improves traceability and stage-level evaluation. After refinement, scenarios contain 4--6 observable stages (mean 5.1) out of 6--7 semantic stages (Table~\ref{tab:stage-coverage}).
This gap reflects telemetry visibility rather than detector failure: stages without IOC-backed runtime evidence are not included in reconstruction denominators.

\begin{table*}[!htbp]
\centering
\scriptsize
\caption{Scenario-level event and IOC distribution in SynthChain.}
\label{tab:synthchain-ioc-distribution}
\resizebox{\textwidth}{!}{
\begin{tabular}{lrrrrrrrrrr}
\toprule
Scenario & $E$ & Split & $T_n$ 
& IOC-All & IOC-Pre & IOC-Tail 
& $u_{\text{tail}}$ & $u_{\text{ioc}}$ 
& Line-All & Line-Tail \\
\midrule
sc1 & 10,500  & 7,349  & 3,151  & 35.47\% & 36.52\% & 33.01\% & 1,485 & 16.57\% & 32.21\% & 29.58\% \\
sc2 & 52,807  & 36,964 & 15,843 & 0.85\%  & 1.15\%  & 0.16\%  & 1,898 & 0.68\%  & 0.81\%  & 0.13\% \\
sc3 & 52,223  & 36,556 & 15,667 & 0.78\%  & 0.25\%  & 2.03\%  & 1,983 & 8.77\%  & 0.56\%  & 1.44\% \\
sc4 & 122,972 & 86,080 & 36,892 & 0.15\%  & 0.07\%  & 0.35\%  & 4,485 & 2.19\%  & 0.11\%  & 0.26\% \\
sc5 & 28,581  & 20,006 & 8,575  & 13.23\% & 9.04\%  & 23.01\% & 2,601 & 27.53\% & 12.93\% & 22.66\% \\
sc6 & 29,324  & 20,526 & 8,798  & 12.43\% & 8.68\%  & 21.20\% & 2,283 & 28.25\% & 9.78\%  & 16.37\% \\
sc7 & 33,606  & 23,524 & 10,082 & 3.39\%  & 0.68\%  & 9.70\%  & 1,952 & 15.16\% & 2.28\%  & 6.01\% \\
\bottomrule
\end{tabular}
}
\end{table*}

Table~\ref{tab:synthchain-ioc-distribution} further reports the resulting event and IOC distribution. IOC-All, IOC-Pre, and IOC-Tail denote the prevalence of $y_i^{ioc}=1$ in the full stream, training prefix, and evaluation tail. 
Line-All and Line-Tail report the stricter line-level prevalence. 
The large variation across scenarios motivates per-scenario reporting in addition to macro averages. The columns $u_{\text{tail}}$ and $u_{\text{ioc}}$ summarize tail diversity after deduplicating repeated interaction patterns. We define an interaction pattern as $(r,u,v)$, consisting of relation type, source entity, and destination entity. $u_{\text{tail}}$ is the number of unique patterns in the evaluation tail, and $u_{\text{ioc}}$ is the fraction of these unique patterns that contain at least one IOC-bearing event. These values indicate whether tail positives are distributed across distinct behaviors or concentrated in repeated event patterns.

\paragraph{Evaluation Metrics}

The evaluated metrics include AUROC, AUPRC, Precision@K, IOC Hit@K, Stage Recall@K and PredStage@K. While AUROC, AUPRC, Precision@K, and IOC Hit@K assess event-level anomaly ranking performance. Stage Recall@K evaluates attack-chain reconstruction by measuring the fraction of observable attack stages covered by at least one event among the top-(K) retrieved interactions. Let $S_{\text{obs}}$ denote the set of observable GT stages and $S_K$ denote the set of stages represented by the retrieved top-(K) events:
\[
\text{Stage Recall@K} =
\frac{|S_K \cap S_{\text{obs}}|}
{|S_{\text{obs}}|}.
\]
We report two variants depending on how $S_K$ is obtained. In the IOC-mapped variant, stages are assigned using GT IOC-stage mappings, which isolates anomaly ranking and evidence retrieval quality. In the deployable predicted-stage variant, denoted as PredStage@K, stages are assigned by the downstream stage decoder, evaluating reconstruction when GT stage annotations are unavailable during operation.

\subsection{Baseline Comparison}

This subsection addresses RQ2 by comparing FuseChain with source-restricted, statistical/event-pattern, and static graph baselines under the same chronological split.
A direct comparison with full APT provenance systems is difficult to interpret in our setting. Systems such as KAIROS~\cite{Cheng2023KairosPI}, ORTHRUS~\cite{jian2025}, MAGIC~\cite{294545}, PROGRAPHER~\cite{287131}, and NODLINK~\cite{li2024nodlink} are primarily designed for enterprise APT detection over host-centric audit provenance, often assuming benchmark-specific provenance schema, attack windows, or campaign-level labels. Runtime SSC detection differs in both input and objective: evidence is sparse, stage-dependent, and distributed across multi-source runtime telemetry sources. Porting these systems would require substantial changes to their input schema, supervision assumptions, and reconstruction objectives, making the resulting comparison confounded by system adaptation choices.

We therefore construct controlled baselines that isolate the main capabilities required for runtime SSC defense. Source-restricted baselines test whether a single telemetry view is sufficient: audit-only approximates host-provenance visibility, network-sensor-only approximates flow-level network detection, and alert-only approximates IDS-alert-centric analysis.

For full-telemetry baselines, we separate statistical/event-pattern methods from static graph models. Freq-rarity and Path-LOF (path-based Local Outlier Factor) score anomalous interaction patterns over the unified event stream without temporal graph learning, while GraphSAGE and RGCN use the same multi-source graph abstraction but omit temporal memory and event-time evolution. This grouping isolates multi-source visibility from temporal graph modeling.

All methods use the same chronological train/test split as FuseChain within each evaluated scenario. Macro averages are computed over the scenarios to which each method applies, as indicated by the Scenarios column in Table~\ref{tab:baseline-benchmark}.

\begin{table*}[!htbp]
\centering
\small
\caption{Detection, reconstruction, and inference latency on SynthChain. \textit{Scen.} reports the number of scenarios included in each macro-average.}
\label{tab:baseline-benchmark}
\begin{tabular}{lcccccc}
\hline
\textbf{Method} & \textbf{Scenarios} & \textbf{AUROC} & \textbf{AUPRC} & \textbf{P@500} & \textbf{StageRec@500} & \textbf{Latency (ms/1k)} \tabularnewline
\hline

\multicolumn{7}{l}{\textit{Single-source baselines}} \tabularnewline
Audit-only (provenance~\cite{Cheng2023KairosPI,10.5555/3766078.3766446,294545}-style) & 6/7 & 0.521 & 0.140 & 0.064 & 0.222 & 37.5 \tabularnewline
Network-sensor-only (network~\cite{9789878}-style) & 3/7 & 0.550 & 0.141 & 0.113 & 0.361 & 35.7 \tabularnewline
Alert-only (IDS)~\cite{10329144}-style & 3/7 & \textbf{0.948} & \textbf{0.418} & 0.127 & 0.222 & 36.4 \tabularnewline
\hline

\multicolumn{7}{l}{\textit{Full multi-source statistical/event-pattern baselines}} \tabularnewline
Freq-rarity~\cite{manzoor2016fastmemoryefficientanomalydetection}-style & 7/7 & 0.553 & 0.145 & 0.055 & 0.226 & 44.6 \tabularnewline
Path-LOF (path-based Local Outlier Factor~\cite{10.1145/342009.335388}) & 7/7 & 0.553 & 0.145 & 0.055 & 0.226 & 34.2 \tabularnewline
\hline

\multicolumn{7}{l}{\textit{Full multi-source static graph baselines}} \tabularnewline
GraphSAGE (static homogeneous graph~\cite{9899459}) & 7/7 & 0.542 & 0.151 & 0.204 & 0.345 & 0.3 \tabularnewline
RGCN (static heterogeneous graph~\cite{tan2026operationalruntimebehaviormining}) & 7/7 & 0.541 & 0.135 & 0.098 & 0.357 & 1.2 \tabularnewline
\hline

\multicolumn{7}{l}{\textit{Full multi-source temporal graph model}} \tabularnewline
\textbf{FuseChain} & 7/7 & 0.638 & 0.198 & \textbf{0.215} & \textbf{0.405} & 176.0 \tabularnewline
\hline
\end{tabular}
\end{table*}

Table~\ref{tab:baseline-benchmark} shows that single-source visibility is insufficient for full runtime SSC defense. The alert-only baseline achieves high AUROC and AUPRC, but only on the IDS-visible subset covering 3 out of 7 scenarios. This indicates that IDS alerts can be highly separable when attacks trigger signatures, but they do not provide consistent coverage across multi-stage SSC scenarios. Audit-only and network-sensor-only baselines similarly miss evidence that appears outside their respective telemetry views.

Among methods evaluated on all seven scenarios, FuseChain achieves the best macro-averaged AUROC, AUPRC, P@500, and StageRec@500. Its improvement over statistical/event-pattern baselines shows that simple frequency or local-density scoring is insufficient for sparse SSC evidence, while its improvement over static graph baselines shows that multi-source graph construction alone does not capture the event-time dynamics needed for attack-stage reconstruction. FuseChain incurs higher latency because temporal memory updates and relation-aware scoring are performed per event, but the cost remains at the millisecond scale per 1k edges.

We do not include LLM-based detectors in the main table. Our protocol scores edge-level anomalies on a canonical temporal graph under weak rule supervision, with latency measured per 1k edges. LLM approaches typically operate on raw log text with different context windows, supervision, and cost profiles, and are not directly comparable without a separate prompting and calibration pipeline. We leave LLM-augmented threat hunting as future work.

\subsection{Detection and Localization}

This subsection addresses RQ1 by evaluating whether FuseChain ranks IOC-containing runtime events above benign activity and exposes suspicious evidence for analyst-facing localization.
Table~\ref{tab:detection-localization} reports event-level precision, IOC recall, and observable stage recall among the top-500 ranked events for each scenario.

\begin{table}[!htbp]
\centering
\caption{Per-scenario self-supervised detection/localization.}
\label{tab:detection-localization}
\begin{tabular}{lrrrr}
\toprule
Sc. & P@500 & IOC Hits & IOC Rec. & Stage Rec. \\
\midrule
sc1 & 0.238 & 119 & 0.114 & 0.500 \\
sc2 & 0.008 & 4   & 0.154 & 0.667 \\
sc3 & 0.030 & 15  & 0.047 & 0.750 \\
sc4 & 0.016 & 8   & 0.062 & 0.167 \\
sc5 & 0.694 & 347 & 0.176 & 1.000 \\
sc6 & 0.108 & 54  & 0.029 & 0.500 \\
sc7 & 0.242 & 121 & 0.124 & 0.167 \\
\midrule
\textbf{Mean} & \textbf{0.191} & \textbf{95.4} & \textbf{0.101} & \textbf{0.524} \\
\bottomrule
\end{tabular}
\end{table}

The results show that FuseChain can recover IOC-containing evidence under highly imbalanced runtime SSC telemetry. On average, the top-500 events contain 95.4 IOC-containing interactions, with a mean P@500 of 0.191 and Stage Rec. of 0.524. This indicates that top-ranked anomalies often expose stage-level attack context even when IOC hits are sparse.

Performance varies across scenarios due to differences in IOC density and tail-label coverage. For example, sc2 and sc4 have low P@500 because only a small number of IOC-containing events appear in the evaluation tail, while sc5 has denser attack evidence and reaches complete observable-stage recovery. This also shows why event-level precision alone is insufficient: a few diverse IOC hits may cover multiple attack stages, whereas repeated hits from the same behavior may not improve reconstruction. FuseChain therefore aggregates high-scoring events into suspicious provenance graphs using temporal proximity, shared entities, and provenance relationships.

\subsection{Attack-Chain Reconstruction}

This subsection addresses RQ3 and part of RQ4 by evaluating whether retrieved anomalous events cover observable attack stages and how retrieval diversity affects reconstruction quality. Reconstruction is performed by applying different retrieval strategies after anomaly ranking; these strategies change only which ranked events are selected, not the anomaly scores produced by the detector.

\begin{table}[!htbp]
\centering
\caption{Attack-chain reconstruction under different retrieval strategies.}
\label{tab:retrieval-comparison}
\begin{tabular}{lcccc}
\toprule
Retrieval & by\_k & PairDedupe & GroupCap & Adaptive \\
\midrule
Stage Recall@500 & 0.524 & 0.643 & 0.560 & 0.655 \\
\bottomrule
\end{tabular}
\end{table}

Table~\ref{tab:retrieval-comparison} shows that reconstruction quality depends not only on anomaly ranking, but also on evidence diversity. The default by\_k strategy recovers 0.524 Stage Recall@500, while PairDedupe and Adaptive retrieval improve recall to 0.643 and 0.655, respectively. This suggests that high anomaly scores can be concentrated in repeated event patterns, causing the top-K list to contain redundant evidence. By suppressing repeated structures and increasing source or pattern diversity, adaptive retrieval exposes a broader set of observable attack stages without modifying the detector itself.

\subsection{Ablation Study}

This subsection further addresses RQ3 by comparing joint and decoupled stage-decoding strategies for deployable attack-stage interpretation. The comparison focuses on whether stage interpretation should be learned jointly with anomaly detection or trained as a downstream layer after anomaly learning. Joint and balanced training optimize stage prediction together with the anomaly objective, rule supervision uses weak per-scenario stage labels, and frozen decoding trains a lightweight stage decoder on top of a frozen self-supervised anomaly backbone.

\begin{table}[!htbp]
\centering
\caption{Stage-decoding ablation.}
\label{tab:decoder-ablation}
\begin{tabular}{lcccc}
\toprule
Method & PredStage & by\_k & Adaptive & P@500 \\
       & @500      & @500  & @500     &       \\
\midrule
Joint training
& 0.369 & 0.393 & 0.333 & 0.092 \\
Balanced joint training
& 0.345 & 0.429 & 0.345 & 0.096 \\
Rule-based (per-scenario)
& 0.167 & \textbf{0.512} & \textbf{0.560} & 0.175 \\
Frozen decoder
& \textbf{0.881} & 0.476 & 0.488 & \textbf{0.226} \\
\bottomrule
\end{tabular}
\end{table}

Table~\ref{tab:decoder-ablation} shows that frozen decoding provides the strongest deployable stage interpretation. PredStage@500 increases from 0.369 under joint training and 0.345 under balanced training to 0.881 with a frozen backbone. This suggests that sparse and imbalanced stage supervision can interfere with anomaly-centric representation learning when optimized jointly, whereas freezing the self-supervised backbone preserves the learned behavioral structure and lets the decoder act as a downstream interpretation layer.

The table also separates deployable prediction from retrieval quality. PredStage@500 evaluates decoder-predicted stages, whereas by\_k and Adaptive use IOC-stage mappings on retrieved events and therefore measure how well anomaly ranking and retrieval expose stage-covered evidence. Frozen decoding achieves the best PredStage@500 and the highest P@500, indicating that decoupled interpretation improves deployable reconstruction without weakening anomaly localization.

\begin{table}[!htbp]
\centering
\caption{Per-scenario PredStage@500 under stage-decoding ablations.}
\label{tab:decoder-ablation-per-scenario}
\begin{tabular}{lcccc}
\toprule
Scenario & Joint & Balanced & Rule & Frozen \\
\midrule
sc1 & 0.500 & 0.500 & 0.250 & \textbf{1.000} \\
sc2 & 0.167 & 0.333 & 0.167 & \textbf{1.000} \\
sc3 & 0.500 & 0.500 & 0.250 & \textbf{1.000} \\
sc4 & 0.333 & 0.167 & 0.167 & 0.833 \\
sc5 & 0.250 & 0.250 & 0.000 & \textbf{1.000} \\
sc6 & 0.667 & 0.333 & 0.167 & \textbf{1.000} \\
sc7 & 0.167 & 0.333 & 0.167 & 0.333 \\
\midrule
\textbf{Mean} & \textbf{0.369} & \textbf{0.345} & \textbf{0.167} & \textbf{0.881} \\
\bottomrule
\end{tabular}
\end{table}

Table~\ref{tab:decoder-ablation-per-scenario} further shows that the improvement is consistent across most scenarios: frozen decoding recovers all observable stages in five scenarios and substantially improves reconstruction in sc4. The main exception is sc7, where all methods remain limited, suggesting that stage prediction is still sensitive to scenario-specific evidence sparsity and distribution shift. Overall, these results support treating attack-stage reconstruction as a downstream interpretation task built on anomaly-centric temporal graph representations, rather than as a jointly optimized objective.

\subsection{LOSO Generalization}

This subsection addresses RQ4 by testing whether learned representations remain useful under leave-one-scenario-out (LOSO) cross-scenario evaluation. In each LOSO run, FuseChain is trained on six scenarios and evaluated on the held-out scenario tail. This setting is more challenging than the primary per-scenario protocol because attack behaviors, telemetry distributions, IOC density, and observable stages vary substantially across SynthChain scenarios. We therefore use LOSO to assess whether the learned representations transfer across scenario families, rather than as the primary deployment setting.

\begin{table}[!htbp]
\centering
\caption{Per-scenario versus LOSO evaluation (7-scenario macro average).}
\label{tab:loso-summary}
\begin{tabular}{lcccc}
\toprule
Setting & $p@500$ & Stage Recall@500 & AUPRC & AUROC \\
\midrule
Per-scenario & 0.191 & 0.524 & 0.194 & 0.616 \\
LOSO         & 0.545 & 0.645 & 0.589 & 0.636 \\
\bottomrule
\end{tabular}
\end{table}

Table~\ref{tab:loso-summary} shows that FuseChain can retain useful anomaly-ranking and reconstruction signals under cross-scenario training. LOSO achieves higher macro-averaged P@500, Stage Recall@500, AUPRC, and AUROC in this stress test, suggesting that some temporal provenance patterns transfer across SSC scenarios. However, these results should not be interpreted as a strict replacement for the per-scenario setting because the two protocols use different model-selection criteria and are affected by scenario-level differences in IOC density and tail-label coverage. Instead, the takeaway is that FuseChain is not limited to memorizing a single scenario, while reliable deployment still requires evaluation under both local and cross-scenario distributions.

\subsection{Weak-Rule Update Stability}

This subsection addresses RQ4 by evaluating whether a small weak-rule update increases coverage without causing a large detection regression. The update adds an analyst rule for
direct \texttt{setup.py} execution in package-install chains, a behavior that is
common in install-time supply-chain payload execution. 

\begin{table}[!htbp]
\centering
\small
\caption{Weak-rule update ablation}
\label{tab:rules-update-ablation}
\begin{tabular}{lrrrr}
\toprule
\textbf{Scen. set} & \textbf{Events} & $\Delta$RuleHit & $\Delta$RuleHigh & \textbf{New-rule hits} \\
\midrule
sc1--sc7 & 330013 & 104 & 104 & 104 \\
\bottomrule
\end{tabular}

\vspace{0.6em}
\begin{tabular}{lrrrrrr}
\toprule
\textbf{Scen.} & \multicolumn{3}{c}{\textbf{AUPRC}} & \multicolumn{3}{c}{\textbf{P@100}} \\
\cmidrule(lr){2-4}\cmidrule(lr){5-7}
 & Base & Updated & $\Delta$ & Base & Updated & $\Delta$ \\
\midrule
sc2 & 0.003 & 0.002 & -0.001 & 0.000 & 0.010 & +0.010 \\
sc6 & 0.195 & 0.205 & +0.009 & 0.060 & 0.040 & -0.020 \\
\midrule
Mean & 0.099 & 0.103 & +0.004 & 0.030 & 0.025 & -0.005 \\
\bottomrule
\end{tabular}
\end{table}

Table~\ref{tab:rules-update-ablation} separates rule-coverage change from detection-regression checking. The coverage subtable applies the updated rule file to all 330,013 processed SynthChain events across sc1--sc7. Compared with the base rule file, the new \texttt{setup.py} rule produces 104 additional rule-matched events; all 104 are assigned to the high-confidence rule-hit category. Thus, $\Delta$RuleHit and $\Delta$RuleHigh are both 104.

The detection-regression check is run on a smaller subset of scenarios. On sc2 and sc6, the updated rules slightly increase mean AUPRC from 0.099 to 0.103, while mean P@100 changes from 0.030 to 0.025. The per-scenario changes are small and mixed: sc2 gains P@100 with negligible AUPRC change, whereas sc6 gains AUPRC with a small P@100 decrease. Overall, the update increases weak-rule coverage for package-install behavior while preserving detection stability on the tested subset.

\section{Discussion and Limitations}

\subsection{Key Granularity and Noise Control}
FuseChain's effectiveness depends on graph key granularity. Coarse keys may merge unrelated behaviors and obscure attack evidence, while overly fine-grained keys may create sparse, noisy graphs dominated by rare benign values. This trade-off is important for runtime SSC telemetry, where package names, command lines, file paths, domains, and network endpoints can be high-cardinality and transient. FuseChain uses typed entity normalization, source metadata, and event attributes to balance abstraction and traceability, but adaptive key abstraction remains future work.

\subsection{Observable Reconstruction Boundaries}
FuseChain reconstructs observable attack stages rather than complete semantic kill chains. Some semantic stages may not leave IOC-backed evidence in the collected runtime telemetry, so Stage Recall@K and PredStage@K measure coverage over observable stages only. Reconstruction quality is therefore bounded by telemetry availability: missing package traces, process logs, DNS/HTTP metadata, or IDS alerts can make some stages weakly represented or unobservable. FuseChain is best viewed as a fusion layer over complementary runtime sensors, not as a replacement for source-specific detectors.

\subsection{Generalization and Deployable Interpretation}

Our results support decoupling anomaly detection from stage interpretation, but the downstream decoder still depends on the coverage and quality of available stage annotations. Scenarios with sparse or shifted stage evidence may remain difficult. In addition, SynthChain provides controlled multi-source SSC scenarios but does not cover all production environments, logging policies, package ecosystems, or attack behaviors. LOSO evaluation provides an initial stress test for cross-scenario transfer, while broader validation on real-world organizational telemetry remains future work.

\section{Related Work}

\begin{table*}[!htbp]
\centering
\caption{Qualitative capabilities in the multi-stage SSC $\times$ multi-source setting.}
\label{tab:qualitative-comparison}
\vspace{0.3em}
\footnotesize
Legend: \fullbullet{} native support, \halfbullet{} partial support, and \emptybullet{} not explicitly targeted.
\vspace{0.5em}
\small
\begin{tabular}{l l c c c c c c c c c c}
\toprule
Framework & Datasets & C1 & C2 & C3 & C4 & C5 & C6 & C7 & C8 & C9 & C10 \\
\midrule
SIGL~\cite{272316} & install-time audit
& \fullbullet & \emptybullet & \emptybullet & \halfbullet & \emptybullet & \emptybullet & \fullbullet & \emptybullet & benign & \emptybullet \\

KAIROS~\cite{Cheng2023KairosPI} & DARPA TC
& \emptybullet & \emptybullet & \emptybullet & \fullbullet & \fullbullet & \emptybullet & \fullbullet & \fullbullet & benign & \fullbullet \\

ORTHRUS~\cite{10.5555/3766078.3766446} & DARPA TC
& \emptybullet & \emptybullet & \emptybullet & \fullbullet & \fullbullet & \halfbullet & \emptybullet & \fullbullet & benign & \emptybullet \\

NODLINK~\cite{li2024nodlink} & DARPA TC
& \emptybullet & \emptybullet & \emptybullet & \fullbullet & \emptybullet & \fullbullet & \fullbullet & \halfbullet & benign+ & \fullbullet \\

MAGIC~\cite{294545} & DARPA TC; StreamSpot~\cite{10.1145/2939672.2939783}
& \emptybullet & \emptybullet & \emptybullet & \fullbullet & \emptybullet & \emptybullet & \fullbullet & \emptybullet & benign & \emptybullet \\

PROGRAPHER~\cite{287131} & DARPA TC; StreamSpot
& \emptybullet & \emptybullet & \emptybullet & \fullbullet & \emptybullet & \fullbullet & \emptybullet & \halfbullet & benign & \emptybullet \\

T-Trace~\cite{li_t-trace_2024} & multi-source APTs; DARPA
& \emptybullet & \fullbullet & \fullbullet & \fullbullet & \emptybullet & \halfbullet & \halfbullet & \fullbullet & benign & \fullbullet \\

Reha et al.~\cite{reha2023anomaly} & temporal provenance graphs
& \emptybullet & \emptybullet & \emptybullet & \fullbullet & \fullbullet & \emptybullet & \emptybullet & \emptybullet & benign & \emptybullet \\

Po\v{s}tuvan et al.~\cite{postuvan_learning-based_2024} & synthetic CTDGs
& \emptybullet & \emptybullet & \emptybullet & \emptybullet & \fullbullet & \emptybullet & \emptybullet & \emptybullet & weak & \emptybullet \\

GeneralDyG~\cite{10.1609/aaai.v39i20.35508} & dynamic graph benchmarks
& \emptybullet & \emptybullet & \emptybullet & \emptybullet & \fullbullet & \emptybullet & \fullbullet & \emptybullet & weak & \emptybullet \\

\textbf{FuseGraph} (ours) & SynthChain
& \fullbullet & \fullbullet & \fullbullet & \fullbullet & \fullbullet & \halfbullet & \halfbullet & \fullbullet & weak & \halfbullet \\
\bottomrule
\end{tabular}
\vspace{2pt}
\begin{minipage}{0.98\linewidth}
\footnotesize
\textit{Notes.}
C1--C10: SSC domain, multi-source fusion, cross-source alignment,
multi-stage behavior, edge anomaly, subgraph anomaly, node anomaly,
attack reconstruction, label regime, and online detection.
\end{minipage}
\end{table*}


\subsection{Software Supply-Chain Attack Detection}

Software supply-chain attacks have evolved from dependency confusion, typosquatting, and package squatting to stealthier package poisoning strategies that exploit installation scripts, dynamic payload loading, environment fingerprinting, and post-install execution. Existing metadata- and static-analysis methods can detect known suspicious patterns, but often miss malicious behaviors that are obfuscated, dynamically fetched, or triggered only under runtime conditions. Recent works therefore leverage execution traces and behavior sequences for malicious package detection~\cite{10.1145/3705304,3698900.3699111,10.1145/3691620.3695262}; however, many still aggregate runtime events into package-level or snapshot-level features, losing temporal dependencies among actions. For instance, file access, environment-variable reads, and network connections may appear benign individually, while their ordered chain can indicate credential collection and exfiltration. Runtime-oriented datasets such as QUT-DV25~\cite{mehedi2025qutdv}, which provides install-time and post-install-time traces for 14,271 Python packages with 36 eBPF-based behavioral features, help reduce the data gap but remain largely trace-level. Our work is complementary by preserving the temporal and relational structure of runtime behaviors, enabling detection models to capture how low-level events interact over time during package installation and execution.

\subsection{Security Graph Construction and Detection}

Prior systems construct provenance or attack graphs by extracting entities and relations from audit logs, endpoint telemetry, and heterogeneous system logs, where processes, files, users, command-line tokens, network endpoints, and alerts are commonly normalized into graph triples~\cite{9899459,10.1145/3719027.3765219,ghosh2026endtoendframeworkfunctionalityembeddedprovenance}. Based on such graph representations, graph-based threat detection methods identify suspicious nodes, edges, paths, or subgraphs, ranging from causal attack-path reconstruction to graph neural network-based anomaly detection and recent temporal graph learning over continuous event streams. FuseChain follows this general direction but targets multi-source temporal software supply-chain telemetry, where evidence is distributed across package execution, system behaviour, DNS records, network flows, TLS metadata, and alert logs. Unlike prior approaches that often focus on single-source provenance or reduce interactions to homogeneous graph structures, FuseChain extracts ontology-consistent triples from Azure, Zeek, Suricata, and package/runtime logs, preserves relation heterogeneity and event attributes, and performs event-level anomaly scoring before aggregating high-scoring events into temporal subgraphs for threat explanation.

\begin{table*}[!htbp]
\centering
\caption{Evaluation-setting characteristics across representative frameworks.}
\label{tab:eval-realism}
\vspace{0.3em}
\footnotesize
\setlength{\tabcolsep}{4pt}
\begin{tabular}{@{}l
  >{\raggedright\arraybackslash}p{0.10\linewidth}
  >{\raggedright\arraybackslash}p{0.14\linewidth}
  >{\raggedright\arraybackslash}p{0.13\linewidth}
  >{\raggedright\arraybackslash}p{0.15\linewidth}
  >{\raggedright\arraybackslash}p{0.12\linewidth}
  >{\raggedright\arraybackslash}p{0.13\linewidth}@{}}
\toprule
Setting &
Threat model &
Attack surface &
Telemetry &
Stage semantics &
Label regime &
Typical metrics \\
\midrule

DARPA TC~\cite{edq8-nk52-21}
& Endpoint APT
& Enterprise hosts; lateral movement, C2, exfil; occasional trojanized \emph{software update} on endpoints
& Single-source provenance (CDM/CamFlow)
& ATT\&CK-aligned APT kill chain
& Campaign / oracle-window GT
& AUROC $\approx 1$ \\

SIGL~\cite{272316}
& Install-time SSC
& Trojanized desktop/server installers
& Single-source OS audit
& Install-time execution
& Installer-level malicious/benign labels
& High P/R on installer corpus \\

SynthChain \cite{tan_2026_synthchain}
& Multi-family SSC
& Package registries (PyPI/npm), installers, cloud runtime, containers/ML artifacts
& Multi-source fusion (audit, Zeek, IDS)
& SSC chain (delivery $\rightarrow$ install $\rightarrow$ runtime $\rightarrow$ impact)
& Weak rules + sparse IOC GT
& Moderate AUROC; stage recall@$K$ \\

\bottomrule
\end{tabular}
\vspace{0.3em}
\small
\textit{Note:} DARPA TC does not cover OSS registries, CI/CD pipelines, or container; SynthChain logs victim-side install/runtime telemetry only.
\end{table*}

\subsection{Anomaly Detection in Temporal Graphs}

Temporal graph anomaly detection aims to identify abnormal nodes, links, events, or subgraphs from evolving interactions. Existing studies model continuous-time dynamic graphs (CTDGs) through temporal node or edge representations and detect anomalies based on structural, temporal, or contextual deviations~\cite{reha2023anomaly,postuvan2024learningbased}. Recent surveys and generalized dynamic-graph detectors further show that temporal dependencies and local evolving structures are important for robust anomaly detection across different graph domains~\cite{10.1145/3669906,10.1609/aaai.v39i20.35508}. However, these methods are mostly designed for generic dynamic graphs or single-source provenance telemetry. FuseChain differs by targeting software supply-chain attacks, where evidence is distributed across heterogeneous runtime telemetry sources. It constructs a multi-source heterogeneous temporal graph, preserves relation types and event attributes, and performs event-level anomaly scoring before aggregating high-scoring events into temporal subgraphs for interpretable threat explanation.

\subsection{Comparison and Positioning}
\label{sec:comparison-positioning}

Table~\ref{tab:qualitative-comparison} positions FuseChain against representative provenance-graph and supply-chain-security frameworks under a multi-stage SSC and multi-source setting. 
We focus on whether each framework naively targets a capability, only partially supports it, or does not explicitly address it.

Most graph-provenance-based threat detection systems are evaluated on DARPA Transparent Computing (TC)~\cite{edq8-nk52-21}, an endpoint-centric APT benchmark in which attacks are injected into controlled host provenance streams with comparatively clean benign baselines and oracle attack windows.
While TC has driven important advances in temporal provenance modeling, its evaluation setting differs materially from industrial SSC defense:
(i)~\emph{attack domain mismatch}---TC emphasizes post-compromise lateral movement and persistence on a single host, whereas SSC threats begin at package registries, build pipelines, and install-time execution with sparse, stage-dependent IOCs;
(ii)~\emph{telemetry homogeneity}---TC evaluations typically rely on a single provenance source (e.g., CamFlow/CDM), while production SSC monitoring must fuse install-time audit logs, IDS alerts, and network telemetry with inconsistent entity schema;
(iii)~\emph{label optimism}---many TC studies report AUROC/AUPRC near 1.0 under campaign-level supervision or subgraphs with known attack intervals, which inflates separability relative to deployment, where analysts operate under partial labels, analyst-added weak rules, and extreme class imbalance in streaming edge logs.
Accordingly, we treat near-perfect TC metrics as \emph{upper-bound feasibility} on APT subgraphs, not as evidence of readiness for multi-stage SSC and multi-source detection.

\section{Conclusion}

We presented \textsc{FuseChain}, a runtime detection framework for SSC attacks based on temporal multi-source provenance graphs. FuseChain aligns heterogeneous telemetry sources into typed temporal events, learns anomaly-centric representations from benign-prefix telemetry, and reconstructs observable attack stages through downstream interpretation. Across seven SynthChain SSC attack scenarios, FuseChain outperforms controlled single-source and static multi-source baselines, while frozen-backbone decoding substantially improves deployable stage reconstruction and adaptive retrieval increases observable-stage coverage without modifying the detector. These results show that runtime SSC defense benefits from combining multi-source temporal provenance modeling with decoupled attack-stage interpretation.

\section*{Acknowledgements}

We thank Chongyang Xu from MPI-SWS for insightful discussions and constructive feedback. Zhuoran Tan was partially supported by JUMPSEC Ltd.


\clearpage
\appendices

\end{document}